\documentclass[journal]{IEEEtran}
\usepackage{graphicx}
\usepackage{array}
\usepackage{cmsrb}
\usepackage[OT2,T1]{fontenc}
\begin{document}
\title{Chorba: A novel CRC32 implementation}
\author{Sam~Russell,~\IEEEmembership{BE Network Engineering}}


\maketitle

\begin{abstract}
This paper describes a novel method for efficiently calculating CRC checksums without lookup tables or hardware support for polynomial multiplication. Throughput of CRC32 is increased by 100\% across different platforms compared with the current state of the art. Performance is on par with or exceeds hardware-accelerated solutions on x86\_64 and ARMv8 processors, and these hardware-accelerated solutions see a performance increase of 5-20\% depending on message length. The small number of operations required with this approach could simplify hardware CRC32 implementations.
\end{abstract}

\IEEEpeerreviewmaketitle

\section{Introduction}
\IEEEPARstart{C}{yclic} redundancy checks are used for error detection across a wide range of applications, including network transmission, and data storage. An application generates the CRC checksum of a message and transmits or stores them together. The CRC algorithm can be run by the receiver at a later time and data corruption can be detected if the output doesn't match the original CRC generated by the sender.

A cyclic redundancy check is a type of checksum that defines a generator polynomial $G(x)$, interprets a string of data as a polynomial $M(x)$ over GF(2), and calculates the checksum $M(x)\ mod\ G(x)$. Any errors in the message are highly likely to result in a different checksum, allowing the fidelity of large files to be verified at the cost of a small amount of extra data. CRC32 uses the generator polynomial $G(x) = x^{32} + x^{26} + x^{23} + x^{22} + x^{16} + x^{12} + x^{11} + x^{10} + x^{8} + x^{7} + x^{5} + x^{4} + x^{2} + x + 1$, or 0x104C11DB7 in hexadecimal.

CRC32 in particular is a very popular standard, used in the Ethernet standard, and various compression standards (GZIP, BZIP2, PKZIP), partially due to the popularity of 32-bit processors from the 80386 in 1986 until the introduction of the AMD64 architecture in 2003.

\section{Related work}
The Sarwate Algorithm \cite{sarwate1988} was widely used, and until recently was the algorithm used in the GZIP compression tool. This improves on the naive polynomial division approach by precomputing a 256 entry lookup table, allowing 8 bits to be processed at a time. The 1024 byte table fits easily in L1 cache on modern processors, making the algorithm quite efficient.

The lookup table approach was extended by \cite{kounavis05}, which uses 4 or 8 tables to process 4 or 8 bytes at a time, meaning the previous CRC value only needs to be combined once per 4 or 8 bytes. Taking advantage of the wider data buses (32 or 64 bits) on modern processors allowed more data to be read in a single instruction. This was further improved in \cite{kadatch}, which performs multiple iterations in parallel due to the fact that modern processors have multiple independent arithmetic and memory lookup units that can complete operations while other instructions are being executed. The fastest CRC32 implementations all use this "braiding" technique on general purpose CPUs with no special opcodes for CRC32 or carry-less multiplication.

For CPUs with accelerated carry-less multiplication (e.g. SSE or AVX on Intel architectures, NEON on ARMv8 architectures), \cite{Gopal10} describes an approach for "folding" 64 bits at a time, by multiplying a 64-bit word of data against the CRC of $x^n$ for an n-bit reduction.

For CRC implementations with few terms, \cite{chi18} is an effective algorithm for implementing Barrett Reductions \cite{barrett1986implementing} with exclusive-or and shift operations, a theme that is built upon in this paper. Ultimately a Barrett Reduction is not as efficient as a "fold" using a single multiplication, but this may be an effective implementation to handle the final few bytes and complete the CRC calculation.

\section{Zero Polynomials}
A zero polynomial $Z(x)$ is defined as any polynomial where $Z(x)\ mod\ G(x) = 0$ for a given generator polynomial $G(x)$. Since $M(x) + Z(x)\ mod\ G(x) = M(x) + 0\ mod\ G(x)$, we can use this identity with any zero polynomial $Z(x)$ to iteratively reduce the degree of any message $M(x)$ over a generator polynomial $G(x)$.

The shortest non-trivial zero polynomial is $G(x)$, as $G(x)\ mod\ G(x) = 0$. With a zero polynomial we can find substitutions that allow us to reduce a larger polynomial. For example, we have the identity that $x^{32} = x^{26} + x^{23} + x^{22} + x^{16} + x^{12} + x^{11} + x^{10} + x^{8} + x^{7} + x^{5} + x^{4} + x^{2} + x + 1$. We can use this to reduce any message polynomial $M(x)$ and replace $x^{n+32}$ with $x^{n+26} + x^{n+23} + x^{n+22} + x^{n+16} + x^{n+12} + x^{n+11} + x^{n+10} + x^{n+8} + x^{n+7} + x^{n+5} + x^{n+4} + x^{n+2} + x^{n+1} + x^n$. This is also equivalent to adding $G(x)^{n-32}$ to the message. This is simply an algorithm for polynomial long division, and is what the naive bitwise implementation of CRC32 does.

Table-based approaches \cite{sarwate1988} \cite{kounavis05} \cite{kadatch} pre-calculate the following lookup table $L(x) = x \times G(x)$ over a given range (0-255 for these examples), and make use of the time-memory tradeoff that a lookup table provides.

The folding approach \cite{Gopal10} extends this by taking the CRC32 of large single-term polynomials and making a similar substitution, however instead of shifting by a single bit at a time, we are able to shift by an arbitrary length. For example, $x^{64}\ mod\ G(x)$ is equal to 0x490D678D, and this allows for a shift of 64 bits. This is the same as saying that 0x100000000490D678D is a zero polynomial. We can then use the same process as above, and iteratively replace 0x10000000000000000 with 0x490D678D, peforming this 64 bits at a time using the intrinsic PCLMUL/VMULL opcode provided by the CPU.

Multiplying polynomials is efficient with hardware support, but as demonstrated in \cite{chi18}, attempting to emulate this with exclusive-or and shift operations is much slower, with the operations per byte rising linearly with the number of terms in the zero polynomial. If we could find zero polynomials with low numbers of terms, or find other techniques to minimise binary arithmetic operations, then we could reduce the cost of polynomial multiplication.

\subsection{Extending the generator polynomial}

We can trivially extend the degree of a given zero polynomial $Z(x)$ by taking its square any number of times, as $Z(x) \times Z(x)\ mod\ G(x) = 0$. This is equivalent to doubling the exponent of each term, as all the intermediate terms cancel out under GF(2), e.g. $(x^a + x^b + ...)^2 = x^{2a} + x^{2b} + x^{a+b} + x^{a+b} + ... = x^{2a} + x^{2b}$. In other words, if $x^a + x^b + ... = 0\ mod\ G(x)$ then $x^{2a} + x^{2b} + ... = 0\ mod\ G(x)$.

\begin{figure}
    \centering
    \includegraphics[width=1.0\linewidth]{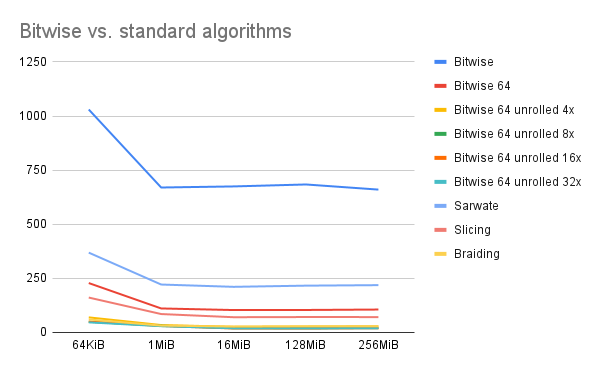}
    \caption{Expanding the generator polynomial with the scaling identity}
    \label{fig:bitwise_standard}
\end{figure}

\begin{figure}
    \centering
    \includegraphics[width=1.0\linewidth]{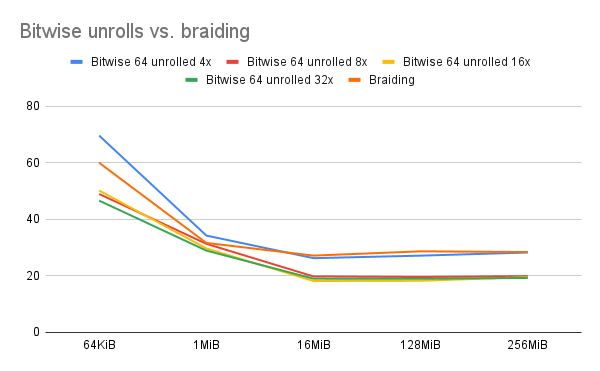}
    \caption{Comparing manually unrolled bitwise loops against braiding}
    \label{fig:bitwise_braiding}
\end{figure}

By using larger zero polynomials we can take advantage of the speed improvements of 8-bit and 64-bit arithmetic and memory accesses. Benchmarking the naive bitwise algorithm against an implementation using a zero polynomial of $G(x)^{64}$ and an 8x unrolled version of this outperformed every other standard implementation. Extending this to a 32x offered no further improvement, and caused regressions on some platforms. These results are displayed in figures \ref{fig:bitwise_standard} and \ref{fig:bitwise_braiding}.

We can further improve performance by choosing zero polynomials with fewer terms, and with other properties which we will discuss later.

\subsection{Finding zero polynomials}
Exhaustively searching for every zero polynomial is a computationally hard problem, but we only need to find polynomials with limited numbers of terms, and of low degree. We can thus precompute $x^n\ mod\ G(x)$ for x from 0 to $2^{20}$ (at the edge of L1 cache and well within L2 cache), and search efficiently inside this space.

$G(x)$ is a primitive polynomial, and so codes repeat every $x^{32}-1$ bits, as described in \cite{peterson1961}. It then follows smallest 2-term zero polynomial is $x^{2^{32}} + 1$ as $x^{2^{32}}\ mod\ G(x) = 1 mod G(x)$. Due to the fact that codes repeat in this fashion, there are no 1-term zero polynomials. There are many zero polynomials of 3, 4 and 5 terms, which we will explore in this paper. Beyond 5 terms it is too costly as we either have too many memory accesses for sparse polynomials, or too many arithmetic operations for dense polynomials.

\subsection{Low degree zero polynomials}

The generator polynomial is the smallest non-trivial zero polynomial of degree 32, with 15 terms. As we reduce the number of terms, the degree of the smallest polynomial increases as follows:

\begin{center}
\begin{tabular}{||c c||} 
 \hline
 Number of terms & Degree \\ [0.5ex] 
 \hline\hline
 15 & 32 \\ 
 \hline
 14 & 42 \\ 
 \hline
 13 & 42 \\ 
 \hline
 12 & 42 \\
 \hline
 11 & 44 \\
 \hline
 10 & 53 \\
 \hline
 9 & 66 \\ 
 \hline
 8 & 89 \\
 \hline
 7 & 123 \\
 \hline
 6 & 203 \\
 \hline
 5 & 300 \\ 
 \hline
 4 & 3006 \\
 \hline
 3 & 91639 \\
 \hline
 2 & $2^{32}-1$ \\
 \hline
 1 & N/A  \\ [1ex] 
 \hline
\end{tabular}
\end{center}

Using the 11-term polynomial of degree 44 resulted in a 50\% performance drop compared with the extended generator polynomial, even when unrolled 8 times. It appeared that we had lost the benefits of data locality, and I decided to focus on polynomials with few terms.

\subsection{Low term zero polynomials}
The polynomial $x^{91639} + x^{49961} + 1$ is the lowest-degree zero polynomial with 3 terms. This performed well in destructive mode (overwriting the data stream in place), but was less performant in nondestructive mode (without modifying the original message), and nondestructive mode is required in most settings. The lowest-degree 4-term zero polynomial is $x^{3006} + x^{791} + x^{140} + 1$. This was passed over in favour of a denser 4-term polynomial which is discussed below.

At 5 terms we seem to find a good tradeoff between number of terms and polynomial degree. $x^{300} + x^{211} + x^{183} + x^{145} + 1$ is the lowest-degree 5-term polynomial. In addition to being the smallest, it is also dense enough to allow the middle 3 terms to be held in memory when calculating in bitwise mode, although with a gap of 66 between terms $x^{211}$ and $x^{145}$ we would expect a performance hit trying to do this with the 64-bit extended variant. I mention this polynomial as it ends up being one of the best performers on the Raspberry Pi 4, which will be discussed in more detail in the Performance section.

\section{Dense zero polynomials}
The polynomial $x^{14870} + x^{22} + x^{11} + x^7 + 1$ is the densest 5-term zero polynomial. This is less useful for bitwise arithmetic, but the scaled polynomial is quite performant. It requires 22 64-bit local variables, which sit inside the register space in ARMv8 systems, and overflow the register space on x64 but are still performant. This requires one read and one write per cycle and zero arithmetic shifts, and substantially fewer exclusive-or operations than the extended generator polynomial.

Also included is the densest 4-term zero polynomial: $x^{5869} + x^{5835} + x^{5821} + 1$. These terms are too far apart for us to efficiently cache them between iterations (48 words, blowing out register space on both x86\_64 and ARM), but this is still quite performant in destructive mode.

Both of these implementations require a final write back for the smallest term, and this incurs a performance penalty compared to the scaled generator polynomial. The lower amount of arithmetic operations due to the lower number of terms makes the tradeoff worthwhile.

\section{Implementation}

For each algorithm we loop over the data in a multiple of 64-bit words, manually execute the polynomial multiplication, and then either write the data ahead into the stream, cache it locally, or put it into a ring buffer. The ring buffer incurs a performance cost as we are now reading from and writing to multiple buffers simultaneously, as well as some arithmetic costs when calculating a buffer address modulo the length of the loop. Choosing a power of 2 for the buffer length offers a substantial improvement in performance as this is calculated with a single bitmask rather than an expensive division instruction.

The algorithms in the graphs are as follows:

\begin{center}
\begin{tabular}{|m{8em} m{14em}|} 
 \hline
 Algorithm name & Implementation \\
 \hline\hline
 braiding & braiding implementation from zlib 1.3.1.1 \cite{gailly2004zlib} \\ 
 \hline
 generator\_64\_bits \_unrolled\_8 & CRC32 generator polynomial expanded 64 bits and unrolled 8 times \\ 
 \hline
 chorba\_352 & $x^{44} + x^{39} + x^{37} + x^{28} + x^{13} + x^{12} + x^9 + x^7 + x^3 + x + 1$ scaled by 8 \\ 
 \hline
 chorba\_small & $x^{300} + x^{211} + x^{183} + x^{145} + 1$ \\ 
 \hline
 chorba\_46952 & $x^{5869} + x^{5835} + x^{5821} + 1$ scaled by 8 \\ 
 \hline
 chorba\_118960 & $x^{14870} + x^{22} + x^{11} + x^7 + 1$ scaled by 8 \\ 
 \hline
 chorba\_733112 & $x^{91639} + x^{49961} + 1$ scaled by 8 \\ 
 \hline
 accelerated (Raspberry Pi 4 and Graviton) & native CRC32 opcode implementation from zlib 1.3.1.1 \cite{gailly2004zlib} \\ 
 \hline
 accelerated (x86\_64) & AVX1-based PCLMUL implementation based on \cite{Gopal10} \\ 
 \hline
\end{tabular}
\end{center}

\subsection{Destructive and non-destructive implementations}

Because we are shifting data so far forward, it is often more efficient to modify the message in place. This is inappropriate in situations where we need to preserve the output (e.g. file compression schemes), but is acceptable in other cases (e.g the cksum tool). With the exception of the short polynomials (chorba\_small, chorba\_352 and generator\_64), both destructive and non-destructive implementations are tested.

\section{Performance}

\begin{figure}
    \centering
    \includegraphics[width=1.0\linewidth]{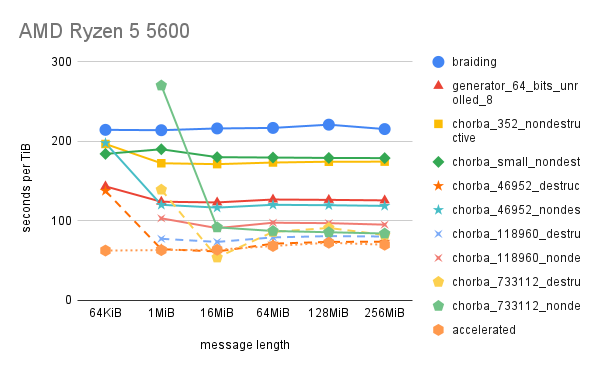}
    \caption{AMD Ryzen 5 5600}
    \label{fig:ryzen}
\end{figure}

\begin{figure}
    \centering
    \includegraphics[width=1.0\linewidth]{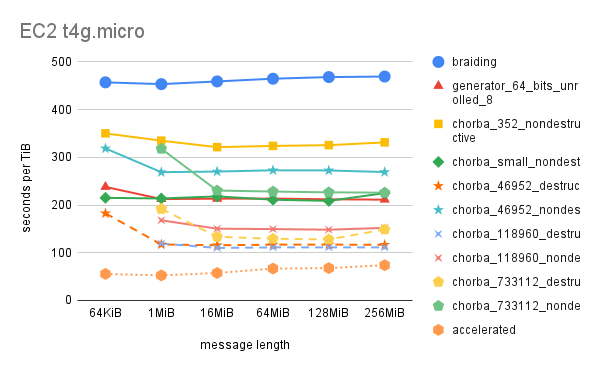}
    \caption{AWS Graviton2}
    \label{fig:t4g}
\end{figure}

\begin{figure}
    \centering
    \includegraphics[width=1.0\linewidth]{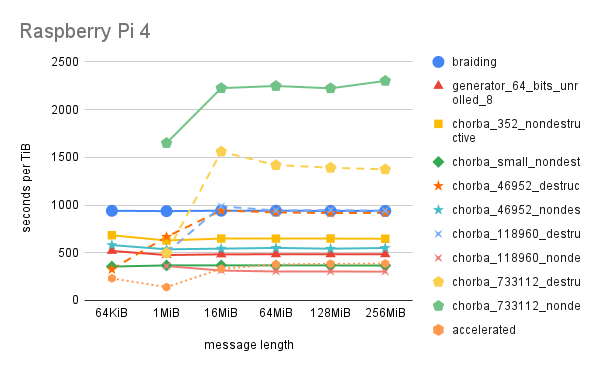}
    \caption{Raspberry Pi 4}
    \label{fig:rpi4}
\end{figure}

The selected polynomials were tested on an AMD Ryzen 5 5600 desktop computer, and an Amazon EC2 t4g.micro instance. Each algorithm was tested with 1000 iterations of messages of length 64KiB, 1MiB, 16MiB, 128MiB, 256MiB and 512MiB. The time recorded was divided by the length of the message and multiplied by 1,000,000,000, giving us a measurement of seconds taken per $10^{12}$ bytes processed.

The results show a significant improvement against braiding, with some subtleties around messages that are near the length of the polynomial being utilised.

One downside to using higher-degree polynomials is a minimum message length where the algorithm can be applied. All Chorba algorithms fall back to chorba\_small\_nondestructive to finish, which is why the throughput is the same at lower message lengths. The non-destructive versions allocate and zero out a memory buffer to begin with, and this ends up costing more than any improvements in throughput for smaller messages. As a result, it makes sense to use the performant polynomials for messages larger than a few megabytes, falling back either to chorba small or the scaled generator polynomial, and to also not bother attempting to initialise the larger polynomials at all for small messages.

\section{Extending AVX implementations}

\begin{figure}
    \centering
    \includegraphics[width=1.0\linewidth]{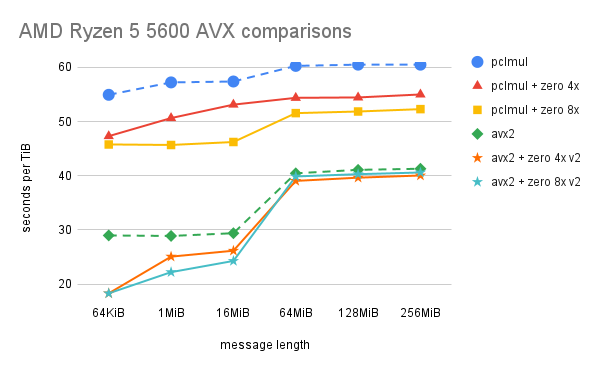}
    \caption{AVX on AMD Ryzen}
    \label{fig:avxryzen}
\end{figure}

\begin{figure}
    \centering
    \includegraphics[width=1.0\linewidth]{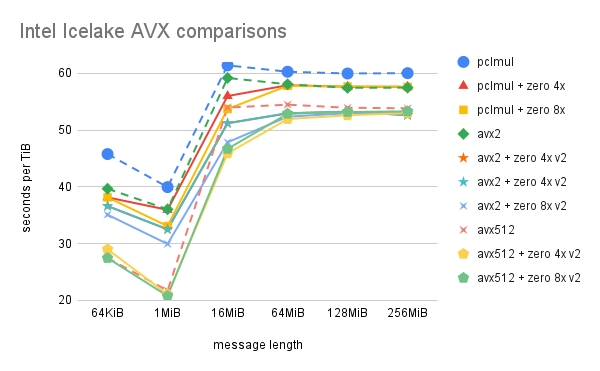}
    \caption{AVX on Intel Skylake}
    \label{fig:avxicelake}
\end{figure}

At the speeds provided by AVX extensions, the bottleneck moves from compute to memory throughput. Implementing Chorba with a scaled $x^{14870} + x^{22} + x^{11} + x^7 + 1$ in AVX1 improves speed slightly, but we see large improvements on AVX1 by interleaving the scaled generator polynomial between PCLMUL executions. AVX2 and AVX512 see similar improvements up to message sizes of 16MiB, beyond which the improvement drops off drastically, presumably due to the message no longer being able to fit in L2 cache.

It is worth noting that on the Intel Icelake processor, while AVX512 outperforms at the 1MiB level, above 16MiB an improved AVX2 implementation outperforms the base AVX512 implementation, and is competitive against an improved AVX512. It appears the "4x" variants are more effective with smaller messages (16MiB and lower), and the "8x" variants are slightly better with longer messages. More work is needed in this area to see whether a different choice of polynomial or other memory access strategies can further improve performance.

The algorithms in the figures are as follows:

\begin{center}
\begin{tabular}{|m{8em} m{14em}|} 
 \hline
 pclmul & AVX1-based PCLMUL implementation based on \cite{Gopal10} \\ 
 \hline
 avx2 & AVX2-based PCLMUL implementation based on \cite{Gopal10} \\ 
 \hline
 avx512 & AVX512-based PCLMUL implementation based on \cite{Gopal10} \\ 
 \hline
 zero 4x & Includes 4 scaled generator polynomial shifts with every 32 folds \\ 
 \hline
 zero 8x & Includes 8 scaled generator polynomial shifts with every 32 folds \\ 
 \hline
\end{tabular}
\end{center}

\section{Conclusion}
CRC32 checksums can be efficiently calculated in software without needing to rely on lookup tables. The use of zero polynomials with a small number of local terms allows not only for efficient software implementations, but also offers possibilities for improving the efficiency of hardware CRC implementations. Non-accelerated implementations approach or even surpass the performance of hardware-accelerated implementations. Hardware-accelerated implementations can be further improved using some of these methods.

Specifically, the polynomial $x^{14870} + x^{22} + x^{11} + x^7 + 1$ (scaled by 8) in non-destructive mode is reliably performant across the systems that were tested, as well as being the outright best performer on the Raspberry Pi 4. This polynomial is likely the best all-around performer and could be deployed as a drop-in replacement for braiding. All 3 destructive polynomials offered superior performance on the AMD and Graviton systems tested, although these should be tailored to individual deployments as the performance regression on the Raspberry Pi 4 suggests that the increased memory writes and wider data working set could become a bottleneck on certain systems.

\section{Future work}
The findings in this paper apply to cyclic redundancy checks of other lengths, and these should be evaluated accordingly. In particular, the CRC-32-XFER and CRC-64-ISO algorithms studied in \cite{chi18} have much shorter generator polynomials (7 and 5 terms respectively), so it is possible that these generator polynomials are the optimal polynomials to use for a Chorba algorithm.

Hardware support using NEON on ARMv8 and AVX on x86\_64 processors improves the speed of polynomial multiplication, but also offers us wider registers and Single Instruction Multiple Data (SIMD) instructions that allow us to parallelise operations. This paper demonstrates an improvement to existing AVX-based implementations and this could potentially improved further.

There are also differences in performance based on whether a polynomial is deployed in a destructive or non-destructive fashion. There may be more efficient ways to manage the intermediate products that are created in non-destructive modes, and these might end up being more efficient than the destructive modes.

\section*{Acknowledgment}

T. Charles Yun, a good friend and mentor, spent a lot of time encouraging me with this research and reviewing earlier versions of the paper. Thank you for your friendship and support over all these years.

Thanks to Jeffrey Walton for allowing me to use his server to benchmark the AVX512 implementations.

\section*{Dedication}

This implementation is named after the Serbian singer Bora Đorđević (also known as Bora Čorba) who was born in 1952 and died in 2024. His birth year matches the number of the GZIP standard RFC 1952 that describes a common CRC32 implementation, and the original proof of concept for this method used the polynomial $x^{21} + x^{15} + x^{14} + x^{11} + x^{10} + x^7 + x^3$
which is $x^{1952 \times 8}\ mod\ G(x)$.

\ifCLASSOPTIONcaptionsoff
  \newpage
\fi

\bibliographystyle{IEEEtran}
\bibliography{bibtex/bib/refs}
\end{document}